\documentclass[twocolappendix,numberedappendix,appendixfloats]{emulateapj}

\usepackage{booktabs} 
\usepackage{float} 
\slugcomment{Accepted for Publication in the Astrophysical Journal}

\shorttitle{HST Proper Motions of the 3C273 Jet}

\shortauthors{Meyer et al.}

\begin{document}


\title{An HST Proper-Motion Study of the Large-scale Jet of 3C273}


\author{Eileen T. Meyer\altaffilmark{1} and William B. Sparks}
\affil{Space Telescope Science Institute, Baltimore, MD 21218}
\email{meyer@stsci.edu}

\author{Markos Georganopoulos\altaffilmark{2}}
\affil{University of Maryland Baltimore County, Baltimore, MD 21250}

\author{Jay Anderson, Roeland van der Marel and John Biretta}
\affil{Space Telescope Science Institute, Baltimore, MD 21210}

\author{Tony Sohn}
\affil{Johns Hopkins University, Baltimore, MD 21210}

\author{Marco Chiaberge}
\affil{Space Telescope Science Institute, Baltimore, MD 21210}

\author{Eric Perlman}
\affil{Florida Institute of Technology, Melbourne, FL 32901}

\author{Colin Norman\altaffilmark{3}}
\affil{Johns Hopkins University, Baltimore, MD 21210}

\altaffiltext{1}{University of Maryland Baltimore County, Baltimore, MD 21250}
\altaffiltext{2}{NASA Goddard Space Flight Center, Greenbelt, MD}
\altaffiltext{3}{Space Telescope Science Institute, Baltimore, MD 21210}

\begin{abstract}
The radio galaxy 3C 273 hosts one of the nearest and best-studied
powerful quasar jets. Having been imaged repeatedly by the Hubble
Space Telescope (HST) over the past twenty years, it was chosen for an
HST program to measure proper motions in the kiloparsec-scale resolved
jets of nearby radio-loud active galaxies. The jet in 3C 273 is highly
relativistic on sub-parsec scales, with apparent proper motions up to
15$c$ observed by VLBI \citep{lister2013}. In contrast, we
  find that the kpc-scale knots are compatible with being stationary,
  with a mean speed of $-$0.2$\pm$0.5$c$ over the whole jet. Assuming
  the knots are packets of moving plasma, an upper limit of $1c$
  implies a bulk Lorentz factor $\Gamma<$2.9. This suggests that the
  jet has either decelerated significantly by the time it reaches the
  kpc scale, or that the knots in the jet are standing shock
  features. The second scenario is incompatible with the inverse
  Compton off the Cosmic Microwave Background (IC/CMB) model for the
  X-ray emission of these knots, which requires the knots to be in
  motion, but IC/CMB is also disfavored in the first scenario due to
  energetic considerations, in agreement with the recent finding of
  \cite{mey14} which ruled out the IC/CMB model for the X-ray emission
  of 3C 273 via gamma-ray upper limits.
\end{abstract}

\section{Introduction}

About 10\% of active galactic nuclei (AGN) produce bipolar jets of
relativistic plasma which can reach scales of tens to hundreds of
kiloparsecs in extent. While there is growing evidence that AGN
feedback, including jet production, may have an important impact on
galaxy and cluster evolution \citep[e.g.][]{fabian2012}, uncertainties about the
physical characteristics of these jets has encumbered attempts to make
these impacts understood quantitatively. Among the chief open
questions are the identity of the radiating particles (positrons,
electrons, or hadronic species), the lifetimes and duty cycles of the
jets, and the magnetic field strength and speed of the plasma. All of
these characteristics feed into the calculation of how much energy
and momentum are carried by these jets and ultimately deposited
into the galaxy and/or cluster-scale environment.  

In theory, proper-motion studies allow us to put direct constraints on the speed
of AGN jets, and consequently their Lorentz factors
($\Gamma$). Hundreds of observations of jets with very long baseline
interferometry (VLBI) in the radio have detected proper motions of
jets on parsec and sub-parsec scales, relatively near to the black
hole engine
\citep[e.g.][]{kellermann1999,giovannini2001,jorstad2001,piner2004,kellermann2004,jorstad2005,lister2009,piner2010}. These
observations show that these jets are often highly relativistic, such
that velocities near the speed of light coupled with relatively small
viewing angles result in apparent superluminal motion. The
dimensionless observed apparent velocity $\beta_\mathrm{app}$ is
related to the real velocity $\beta=v/c$ (where $c$ is the speed of
light) and viewing angle $\theta$ through the well-known Doppler
formula $\beta_\mathrm{app}=\beta \sin \theta/(1 - \beta \cos
\theta)$. A measurement of $\beta_\mathrm{app}$ implies both a lower
limit on the Lorentz factor
($\Gamma_\mathrm{min}\approx\beta_\mathrm{app}$) and an upper limit on
the viewing angle -- constraints which are very difficult to derive
using other means such as spectral fitting, due to the inherent
degeneracy between intrinsic power, angle, and speed introduced by
Doppler boosting of the observed flux.

While proper motions of jets on parsec scales exist in large samples,
direct observations of jet motions on much larger scales (kpc-Mpc) are
rare. Such observations naturally rely on sub-arcsecond resolution
telescopes like the Very Large Array (VLA), Atacama Large
Millimeter/submillimeter Array (ALMA), or Hubble Space Telescope (HST)
in order to image the full jet in detail, but this (much lower than
VLBI) resolution necessarily also limits potential observations of
apparent motions to sources in the very local Universe, and require
years or even decades of repeated observations. Superluminal motions
on kpc scales also might not be common, as the jet presumably
decelerates as it extends out from the host galaxy, though it must be
at least mildly relativistic to explain jet one-sidedness. Statistical
studies of jet-to-counterjet ratios in the radio generally suggest
that kpc-scale quasar jets are only mildly relativistic \citep[$\Gamma\sim
few,$][]{arshakian2004,mullin2009}.

A common problem for both VLBI-scale and kpc-scale
  proper-motion studies is the interpretation of slow-moving or
  stationary features in jets.  While observed motions imply a
  corresponding minimum bulk speed, fast bulk speeds could also be
  present in jets that do not produce convenient ballistic features
  which can be easily tracked, and stationary or slow-moving features
  may instead correspond to stationary shocks within the flow. An
  obvious example is the famous knot HST-1 in the jet of M87, which is
  thought to be a standing recollimation shock, through which plasma
  is moving with a moderately relativistic bulk speed
  \citep{biretta1999,stawarz2006,cheung2007}. A similar feature has
  been seen in the radio in 3C 120 \citep{agudo2012}. On VLBI scales,
  stationary features can also appear in jets along with moving
  components. Some two-thirds of the objects in the VLBI proper motions study of
  BL Lacs by \cite{jorstad2001} contained stationary features, a
  common finding in VLBI studies generally
  \cite[e.g.,][]{alberdi2000,lister2009}.


For many years, there were only two measured proper motions of jets on
kpc scales, both taken with the VLA. These were the famous result of
$\beta_\mathrm{app}$ up to $6c$ measured by \cite{biretta1995} for the
jet in M87 (z=0.004, d=22 Mpc), and a speed of $\approx 4c$ for a knot
in the jet in 3C 120 (z=0.033, d=130 Mpc) by \cite{walker1988}, though
this was later contradicted by additional VLA and Merlin observations
\citep{muxlow1991,walker1997}. In 1999, the first measurement of
proper motions in the optical was accomplished by \cite{biretta1999},
using four years of HST Faint Object Camera (FOC) imaging to confirm
the fast superluminal speeds in the inner jet of M87. However, until
recently \citep{meyer2013_m87} it was unclear if M87 would prove to be the
only superluminal jet on kpc scales.

With the continued development of high-precision astrometry techniques
to align HST
images\footnote{http://www.stsci.edu/$\sim$marel/hstpromo.html}, it is now
possible to register images of jets repeatedly observed by HST over
the past 20 or more years for proper motions studies.  With a single
moderately deep HST image, it is possible to build a reference frame using
background or stationary sources on which to register previous
archival imaging to high precision. In many cases, the longest
baselines are supplied by the early WFPC2 snapshot programs which
targeted bright radio galaxies from 1994 through 1998 (e.g., HST
programs 5476, 5980, 6363). The first successful application of
high-precision HST astrometric methods to jet proper motions was done
for the jet in M87, where we were successful in matching over 400 raw
images of the jet taken from 1995 through 2008 using globular clusters
in the host galaxy \citep[][]{meyer2013_m87}. The \cite{meyer2013_m87}
study greatly improved on previous efforts both in lengthening the
time baseline and in reaching errors on the speed measurements as low
as 0.1$c$, allowing us to measure both transverse motions and
decelerations for the first time.

\begin{figure}[t]
\begin{center}
\includegraphics[width=3.4in]{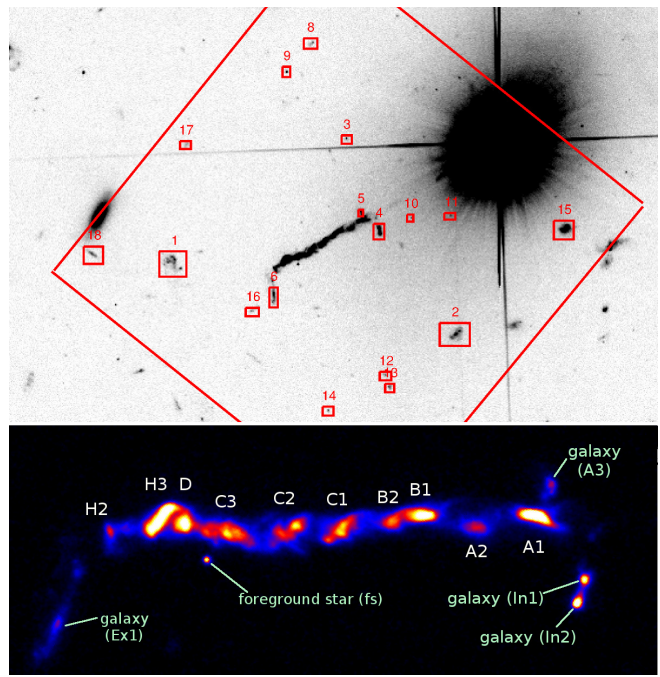}
\caption{
\label{fig:referencesys} {\sl Upper Panel:} The immediate environment of 3C~273 as seen in the ACS/WFC
  reference image from 2014. The intensity has been scaled to
  emphasize the background sources, so the jet appears overexposed
  (center).  The host galaxy/jet core is the extremely bright source
  at right.  The thirteen galaxies used in the registration of the
  1995 epoch are shown, boxed. The larger box outline is roughly the
  field of view for the PC chip in 1995. {\sl Lower Panel:} View of
  the 3C 273 optical jet after galaxy light subtraction.  Knots are
  labeled according to standard convention, as well as several
  nearby background galaxies.}
\end{center}
\end{figure}

In the \cite{meyer2013_m87} study of M87, we reached a limiting astrometric precision in
measuring the positions of knots in the jet of a few mas or less.  Over a
twenty year baseline, this translates into a distance limit of
$\approx$600 Mpc\footnote{Angular-size distances are used throughout
  this paper, with $H_0$=69.6, $\Omega_M$=0.286, $\Omega_\Lambda$=0.714.},
for a target accuracy of 1$c$ in the measurement of superluminal
motion. The handful of optical jets within this local volume which
were first observed in the 1990s are thus ripe targets for HST proper
motions studies. Based on the success of the \cite{meyer2013_m87} study, we
were awarded ACS/WFC observations in cycle 21 for 3 additional nearby
jets previously imaged by HST, including 3C 273, the results of which
are presented here. The results for accompanying target 3C 264, a jet
similar to M87 but 5 times more distant, were published in
\cite{mey15_nature}, while those for 3C~346 will be published
separately. 

At a redshift of 0.158 (d = 567 Mpc), 3C 273 is the furthest kpc-scale
proper-motions target yet attempted with \emph{HST}. The large-scale jet extends
nearly 23$''$ from the core (see Figure~\ref{fig:referencesys}) and
has been observed extensively from radio to X-rays over the past few
decades
\citep[e.g.,][]{schmidt1978,conway1981,tyson1982,lelievre1984,harris1987,thomson1993,bahcall1995,jester2001,marshall2001,sambruna2001,jester2005,jester2006,uchiyama2006,jester2007}. The
X-ray jet of 3C~273 is one of a group of ``anomalous'' X-ray jets
discovered by \emph{Chandra}, where the X-ray emission is too hard and
at too high a level to be consistent with the known radio-optical
synchrotron spectrum \citep{jester2006}. The generally favored model
up until very recently was that these X-rays were produced by inverse
Compton upscattering of CMB photons by a jet still highly relativistic
on kpc scales, to match high speeds implied by parsec-scale VLBI
proper motions \citep{tavecchio2000,celotti2001}. As we discuss in
this paper, this model implies that the knots in jets like 3C 273
should move with significant proper motions. However, the IC/CMB model
was recently ruled out based on gamma-ray upper limits \citep{mey14},
a method first described in \cite{geo06}, and the strong upper limits
placed by our proper motion observations as described in this paper also strongly
disfavor an IC/CMB origin for the X-rays in 3C~273.

\begin{table}[!t]
\caption{HST Imaging Data} \centering
\label{tabledata}
\begin{tabular}{lllcccc}
\toprule
Epoch & PID            & Instrument & Date              & Filter & Exp.           & No.\\
      &                &            & {\tiny (mm/year)} &        & {\tiny (s)}    &    \\
\midrule
1995 & 5980\phantom{7} & WFPC2/PC   & 06/1995           & F622W  & 2300           & 1 \\
     &                 &            &                   &        & 2500           & 1 \\
     &                 &            &                   &        & 2600           & 2

 \\
2003 & 9069\phantom{7} & WFPC2/PC   & 04/2003           & F622W  & 1100           & 2 \\
     &                 &            &                   &        & 1300           & 6 \\
2014 & 13327           & ACS/WFC    & 05/2014           & F606W  & \phantom{0}550 & 4 \\
     &                 &            &                   &        & \phantom{0}598 & 12 \\
\bottomrule
\end{tabular}
\vspace{20pt}
\end{table}

The paper is organized as follows: in Section~\ref{sec:methods} we
present our methods including use of background galaxies to register
the images; in Section~\ref{sec:results} we present the resulting
proper-motion limits, and in Section~\ref{sec:discussion} we discuss
the implications that slow speeds have for our understanding of the
physical conditions in the outer 3C~273 jet.  In Section~\ref{sec:conclusion} we
summarize our conclusions.

\section{methods}
\label{sec:methods}
The data used for this project is summarized in Table~\ref{tabledata},
where we list the project number, instrument setup, date of
observation, and exposure time(s) for the individual exposures,
organized into epochs. Only the 1995 imaging has been previously
reported in \cite{jester2001}. We limited the study to observations in
`V-band' F606W or F622W filters (noting that the wavelength range of
the latter is entirely within the range of the former) for
consistency.

\begin{figure*}[!t]
\begin{center}
\vspace{10pt}
\includegraphics[width=5in]{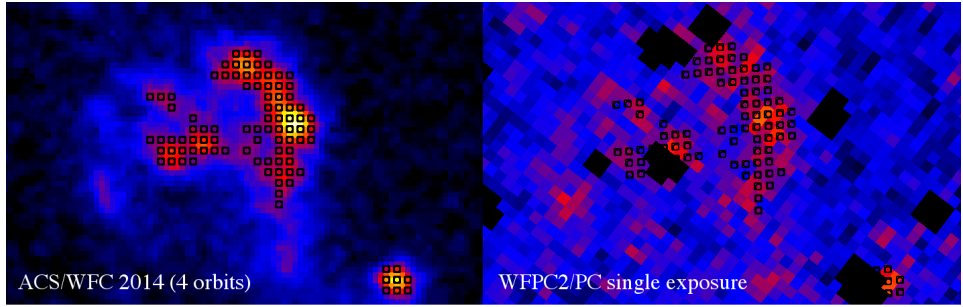}
\caption{\label{fig:sampling} At left, galaxy No. 2 from
  Figure~\ref{fig:referencesys} as seen in the ACS reference image.
  The black points are a portion of the `sampling grid' described in
  the text, which sample the flux distribution of the galaxy. At
  right, the same galaxy in a single exposure (jca201l2q) from the
  1995 WFPC2/PC set. The reference grid has been transformed based on
  an initial solution into the coordinates of the raw exposure, as
  shown. A flux-level cut was applied to only select the grid
  points falling on the brightest parts of the image, so that the
  positions can be compared more easily.  Note that the black areas in
  the right image are locations of cosmic ray artifacts, where flux is
  set to equal zero to make the image more clear.}
\end{center}
\end{figure*}

Before analyzing the archival WFPC2 data, each raw image was
separately corrected for CTE losses.  These losses are increasingly
significant in WFPC2 data with time; we found that without making such
a correction the fluxes measured in the jet and background galaxies were underestimated by
$\approx$5 and $\approx15$\% in the 1995 and 2003 stacks, compared
with the ACS deep stack.  The correction was done pixel-by-pixel since
the jet is resolved and losses depend on the x and y location on the
detector; the method of calculating the correction maps is described
in Appendix~\ref{appendixA}. Note that the CTE correction was not
found to have any effect on image registration or measurement of proper
motions.


\subsection{Reference Image}
The four orbits of ACS/WFC imaging obtained in May of 2014 were
stacked into a mean reference image (with cosmic-ray rejection) on a
super-sampled scale with 0.025$''$ pixels. The registration of the 8
individual exposures utilized a full 6-parameter linear transformation
based on the distortion-corrected positions of 15-20 point-like
sources. The median one-dimensional rms residual relative to the mean
position was 0.07 reference-frame pixels, or 1.75 mas, corresponding
to a systematic error on the registration ($\times 1/\sqrt 16$) of
0.44 mas, or about 2 hundredths of a pixel.

%

The final science image was scaled to monochromatic flux at 6000
\AA$\,$, where the PHOTFLAM value was recalculated in IRAF/STSDAS
package {\tt calcphot} with a power-law model, $\nu^{-1}$, in keeping
with the spectral index reported in \cite{jester2001}. We
also included a reddening/extinction correction with E(B-V)=0.018 for
the position of 3C 273 as derived from the publicly available online
DUST tool\footnote{http://irsa.ipac.caltech.edu/applications/DUST}.

\subsection{Background Source Registration}
To create the 1995 and 2003 epoch science images, an astrometric
solution was found between each individual
(geometrically-corrected) exposure and the reference frame based on
the 2014 ACS image.  Typically, this is accomplished by identifying
background point sources in the deep reference image which constitute
the reference frame of sources used to register the prior epochs.
While some globular clusters associated with the 3C 273 host galaxy
can be seen in the deep ACS image, these are not detected in the much
noisier PC imaging. Instead, we identified 18 background galaxies
based on the criteria that they can be seen by eye above the noise in
the individual PC exposures.  These reference galaxies are highlighted
in Figure~\ref{fig:referencesys}. Note that galaxies 4, 5, and 6
have been previously identified as unrelated to the jet by their lack
of radio emission, and the bright point source near the jet is
actually a foreground star (and thus unsuitable for registering
images due to likely proper motions).

To match the archival images to a common reference frame, our general
strategy was to use the shape and light distribution of the galaxy to
assist in matching their locations in each image. Instead of
identifying a single location associated with each galaxy in the deep
reference image, we instead sample the galaxy in a grid pattern,
resulting in a list of positions along with the flux at each point,
sampling across each galaxy. For example, we show at left in
Figure~\ref{fig:sampling} one of the background galaxies from the
deep 2014 image. Grid points are placed at pixel centers, where we
show only those where the pixel value is $>$10 times the background
for illustration.  At right, we have used an initial transformation
solution to map the points to locations on the galaxy in a single raw
frame.  The flux at each point is interpolated based on nearby pixels.

We used the geometric correction solutions to first create
geometrically-corrected (GC) images from the individual exposures.  An
initial (astrometric) transformation solution was found by supplying
$\approx$10 pairs of matched locations found by hand between the GC
image and the ACS reference image and calculating the six
transformation parameters (without match evaluation/rejection). This
initial transformation solution was then used to create a `rough' mean
image stack for each epoch (in counts units). This stacked image was
then `reverse-transformed' to create a reference image on the scale of
each individual (distorted) raw exposure, in order to detect cosmic rays. These
were detected by initially looking for pixels at a high (10$\sigma$)
deviation from the reference image value, and then masking all
adjacent pixels until all surrounding pixels are near to the mean
value for that pixel. A mask for all pixels flagged as cosmic rays (as
well as for a bad row at x=339 in the 1995 exposures) was thus created
for each raw exposure.

\subsection{Optimizing the Transformation}
The initial transformation solution described above is used as a
starting point to transform the x,y locations for each galaxy grid in
the reference frame into $x_{gc},y_{gc}$ location in the geometrically
corrected image as shown in and discussed previously for
Figure~\ref{fig:sampling}. The intensity can then be sampled at each
location in the GC image, to be compared directly to the scaled counts
value predicted by the scaled reference pixel value.  For each galaxy
in each individual (GC) exposure, we shift the $x_{gc},y_{gc}$ over a
grid of $\delta$x, $\delta$y values, in steps of 1/10th of a pixel for
a total testing range of $\pm$2 pixels. At each point in the grid, a
`score' equal to the sum of squared differences is calculated for the
sampling grid (dropping points falling on cosmic ray artifacts) based
on the updated $x_{gc},y_{gc}$ positions in the GC image. 
We then fit the score matrix with a 2-dimensional Gaussian using IDL
routine 2DGAUSSFIT in order to find the value of $\delta$x, $\delta$y
corresponding to a globally consistent minimum which corresponds to
the optimal position shift which is used to update the location of the
galaxy in the reference frame.

For each individual exposure, we then compile an updated list of
position matches between the reference frame and the GC image from the
mean x,y value of the galaxy sampling grid in each. In general, we used a subset
of the background galaxies which were identifiable by eye and not
overly affected by cosmic ray hits. The process of finding the initial
$x_{gc}, y_{gc}$ values, followed by finding the optimal
$\delta$x,$\delta$y improvement on the mean position, was iterated
until the positions of galaxies stopped improving.

\begin{figure*}[t]
\begin{center}
\includegraphics[width=5.25in]{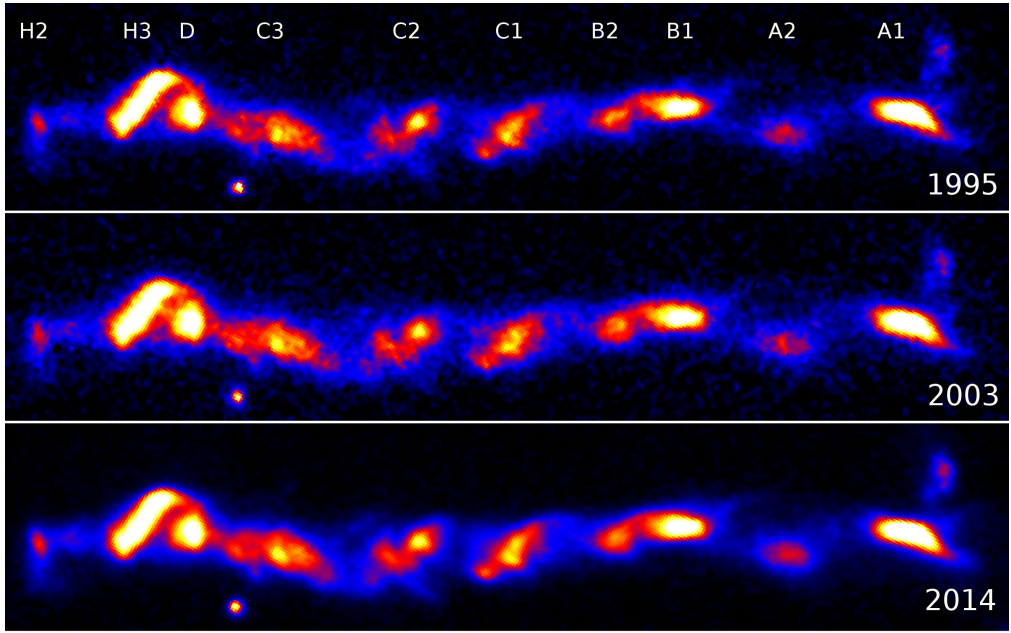}
\caption{\label{fig:3jet} The jet of 3C 273 as seen in 1995, 2003, and
  2014.  The first two images were taken with the WFPC2/PC, the most
  recent with the ACS/WFC.  All have been scaled to monochromatic flux
  at 6000 Angstroms (white = 0.01 $\mu Jy$) after CTE correction, and
  background and host galaxy light subtraction. As shown, no changes
  in the jet are readily apparent by eye.  The bright foreground star
  below knots D and C3 does appear to move with a speed of about 2
  mas/year. Note that we follow the knot labeling of \citep{marshall2001}.}
\end{center}
\end{figure*}

The final science image stacks at each epoch have been scaled to
monochromatic flux at 6000 $\AA$, with background and host galaxy
light subtracted, and are shown in Figure~\ref{fig:3jet}. Note that we
use the knot labeling originally defined in \cite{marshall2001} and
not the later, different labeling used in \cite{uchiyama2006} and
\cite{jester2007}.

\subsection{Measuring Speeds}

We employed two methods to measure the positions of the 10 individual
knots, as well as the 4 background galaxies and foreground star
identified in Figure~\ref{fig:referencesys}, in each of the three
epochs. First, similar to the methods employed in the studies of M87
and 3C~264 \citep{meyer2013_m87,mey15_nature}, we used a centroid
position (flux-weighted mean $x$ and $y$ location) inside a contour
surrounding the brightest part of the knot (hereafter referred to as
the `contour method').  For the brighter knots (A1, B1, D and H3), we
used the 50\% peak flux-over background contour as measured using a
cosine-transform representation of the image. For the fainter knots
where the flux is only slightly higher than the local background
(making it difficult to form a closed contour), we simply used a fixed
circular aperture, centered on the brightest part of the knot (radius
of the aperture depends on the size of the feature and is given in
Table~\ref{table:results}).

As a consistency check, we also measured the shift of each knot using
a second method which we refer to as the `cross-correlation'
method. Over a grid with sub-pixel spacing of 0.2 super-sampled pixels
(5 mas), we shifted the 1995 and 2003 images of each individual knot
relative to the 2014 image (same cutout area) over a 6x6 pixel area,
evaluating the sum of the squared differences between interpolated
flux over the knot area for each x/y shift combination. The resulting
sum-of-squared differences image in all cases clearly shows a smooth
`depression' feature which is reasonably well-fit by a two-dimensional
Gaussian under the transformation $g = 1 - f/\mathrm{max}(f)$, where
$f$ is the original sum of squared differences. Taking the minimum $f$
location as measured by the peak of the two-dimensional Gaussian fit,
we measure the optimal shift for each knot.

To measure the approximate error on the positions measured, we
repeated both of the above methods for simulated images of the jet at
each epoch. The simulated images were created by taking the deep 2014
ACS image and adding a Gaussian noise component appropriately scaled
from the counts in the original WFPC2 exposures. Since the 2014 image
itself has some noise, and also a slightly different PSF from the
WFPC2 images, this method likely slightly overestimates the errors. We
take the error on each knot measurement to be the standard deviation
of the measurements in the simulated images (10 in each epoch).

Finally, we plotted the position of each feature relative to the 2014
position, versus time, to look for evidence of proper motions. We have
transformed from the coordinate frame of the aligned images (North up)
to one based on the jet, where positive $x$ is in the outflow
direction along the jet (taken as position angle (PA) 42$^\circ$ south of
east) and positive $y$ is orthogonal and to the north of the jet. The
data are listed in table \ref{table:results} and plotted in
Figures~\ref{fig:speeds1} and \ref{fig:speeds2}. For both methods, the estimated
error on the measurement has been convolved with the systematic error
of the registration, which is 0.18, 0.22, and 0.02 reference pixels
(4.5, 2.8, and 0.5 mas) for the 1995, 2003, and 2014 epochs,
respectively.


\section{Results}
\label{sec:results}

We first show in Figure~\ref{fig:3jet} a comparison of the jet of 3C
273 in each epoch, where the background and host galaxy light has been
subtracted, and the jet rotated to horizontal. No obvious changes in
the jet are discernible by eye, and the fluxes of all components (as
well as background sources) are consistent to within 5\%. The only
moving component, easily seen when blinking the 1995 and 2014 images
against one another, is the foreground star near knot C3, which
exhibits an apparent motion of 1.9 mas/year, at an angle of
20$^\circ$ north of the the jet direction (where the jet PA is
222$^\circ$). The proper motion is typical for disk stars in our own
galaxy \citep[e.g.,][]{deason2013}, and the star has a V-band
magnitude of 25.6 (STMAG system) and color $m_{F606W}-m_{F814W}$ =
2.9, consistent with the source being a milky way foreground star.

\begin{figure*}[t]
\centering
\includegraphics[width=6in]{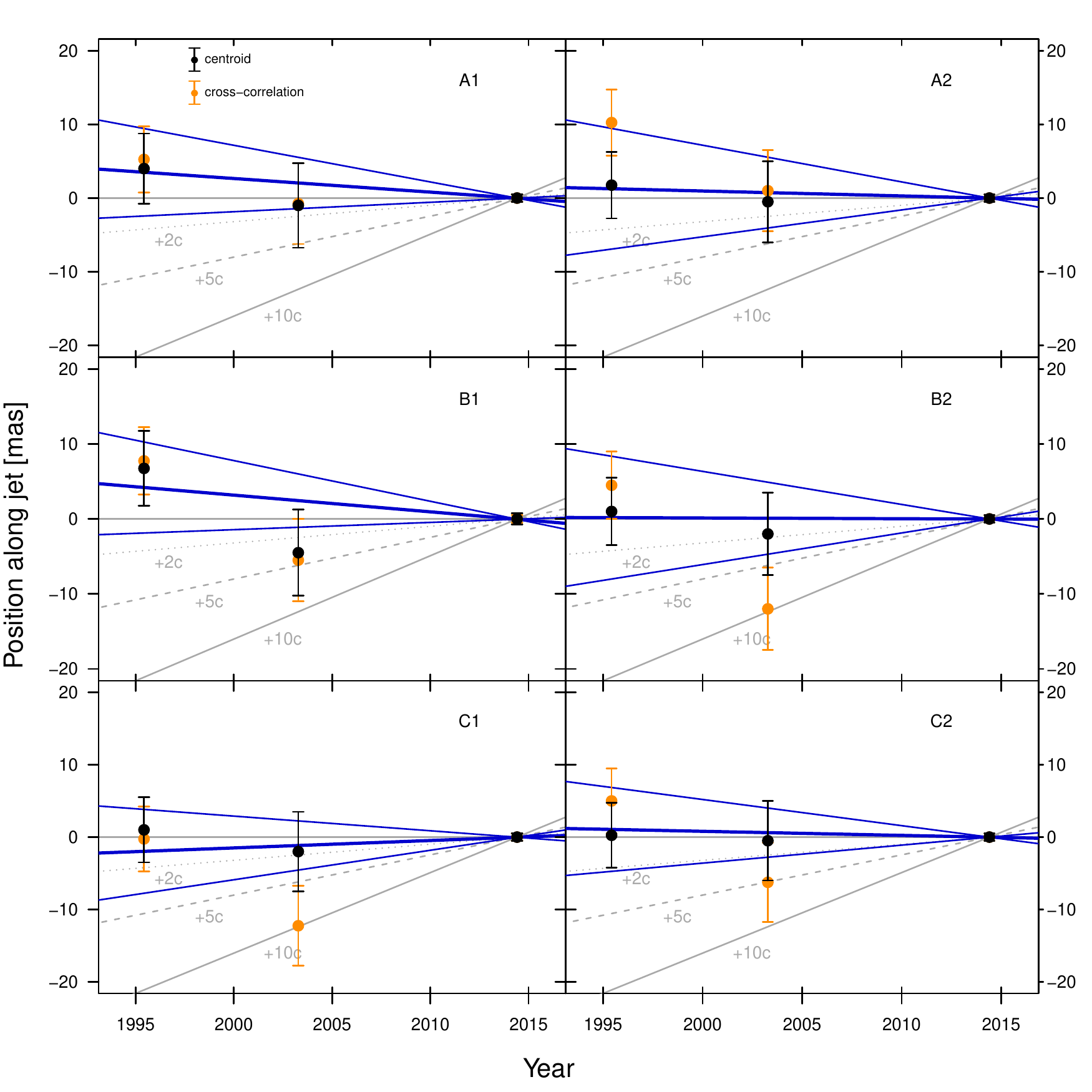}
\caption{\label{fig:speeds1} Shift of individual knots (name noted at
  upper right of each panel) versus time relative to the 2014 measured
  position.  Black points represent the contour-derived shifts and
  orange points the cross-correlation derived shifts. As a guide,
  lines corresponding to an apparent forward speed of 2$c$ (dotted
  gray) and 5$c$ (dashed gray) and 10$c$ (solid gray) are plotted in
  each subfigure. The thick solid blue line is the best-fit weighted
  linear regression model to all points, while the thinner blue lines
  show the 2$\sigma$ (95\%) upper and lower limit slopes. }
\end{figure*}

\begin{figure*}
\centering
\includegraphics[width=6in]{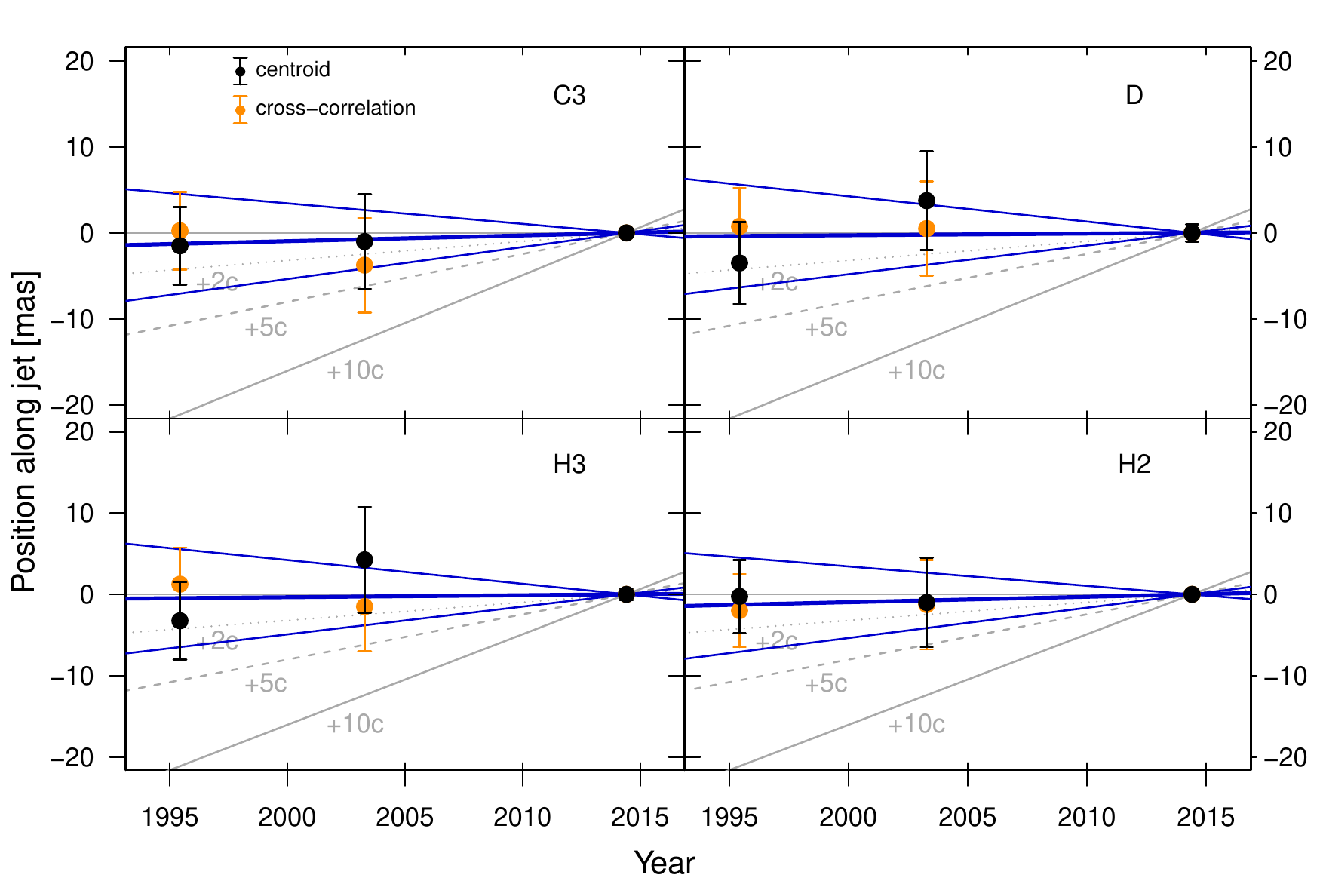}
\caption{\label{fig:speeds2} Shift of individual knots (name noted at
  upper right of each panel) versus time relative to the 2014 measured
  position.  Black points represent the contour-derived shifts and
  orange points the cross-correlation derived shifts. As a guide,
  lines corresponding to an apparent forward speed of 2$c$ (dotted
  gray) and 5$c$ (dashed gray) and 10$c$ (solid gray) are plotted in
  each subfigure. The thick solid blue line is the best-fit weighted
  linear regression model to all points, while the thinner blue lines
  show the 2$\sigma$ (95\%) upper and lower limit slopes.}
\end{figure*}

In Figures~\ref{fig:speeds1}~and~\ref{fig:speeds2}, we have plotted
the shift of each knot relative to 2014 versus time, where black
points represent the contour-derived shifts and orange points the
cross-correlation derived shifts. As a guide, lines corresponding to
an apparent forward speed of 2$c$ (dotted gray) and 5$c$ (dashed gray)
and 10$c$ (solid gray) are plotted in each subfigure. The thick solid
blue line is the best-fit weighted linear regression model to all
points, while the thinner blue lines show the 2$\sigma$ (95\%) upper
and lower limit slopes.  While the two methods of measuring shifts
agree well for most knots, the deviation of the cross-correlation
method from the contour method increases with decreasing surface
brightness.  We have thus excluded the cross-correlation derived
points from the linear fitting for the two knots of particularly low
surface brightness, A2 and B2, though these points are still plotted
in orange in Figure~\ref{fig:speeds1}.

In Table~\ref{tab:speeds} we report the results of the weighted linear
regression fit to the position measurements for the 10 identified
knots in Figure~\ref{fig:referencesys}, as well as four nearby
background galaxies and the foreground star near knot C3. In column 1
we give the knot or object name, in column 2 the measured flux of the
knot in $\mu$Jy, in column 3 the aperture used to measure the surface
brightness given in column 4 in $\mu$Jy/arcsec$^2$. In columns 5 and 6
we give the measured angular speed in mas yr$^{-1}$ in the $x$ (along
the jet) and $y$ (perpendicular to jet) directions, and in columns 7 and
8 the corresponding apparent speeds $\beta_{app,X}$, $\beta_{app,Y}$
in units of $c$, using the conversion factor 8.9856 $c$/(mas
yr$^{-1}$). While these latter values are incorrect/unphysical for the
four galaxies and foreground star (as they are at different/unknown
distances), we include the conversion to $\beta_{app}$ in
Table~\ref{table:speeds} as a convenient reference for the accuracy of
our measurements, since these sources should be completely
stationary. In column 9 we give the probability that the speed of the
knot is greater than zero, and in the final column the 99\% upper limit
on $\beta_{app,X}$.

As shown, all knots have speeds consistent with zero within the errors
of our measurements.  The mean speed along the jet, combining
  all knot values in column 7, is $-0.2\pm0.5c$ As an additional
check, we ran a cross-correlation analysis as described above for
individual knots for the entire optical jet region.  The best-fit
weighted linear regression line yielded a slope of $-$0.006$\pm$0.22
mas yr$^{-1}$ and 0.12$\pm$0.22 mas yr$^{-1}$ along and perpendicular
to the jet, and corresponding to apparent speeds of $-$0.04$\pm$1.9$c$
and 1.1$\pm$1.9$c$, respectively, also consistent with a speed of zero
in both directions.



\begin{figure*}
\centering
\includegraphics[width=6in]{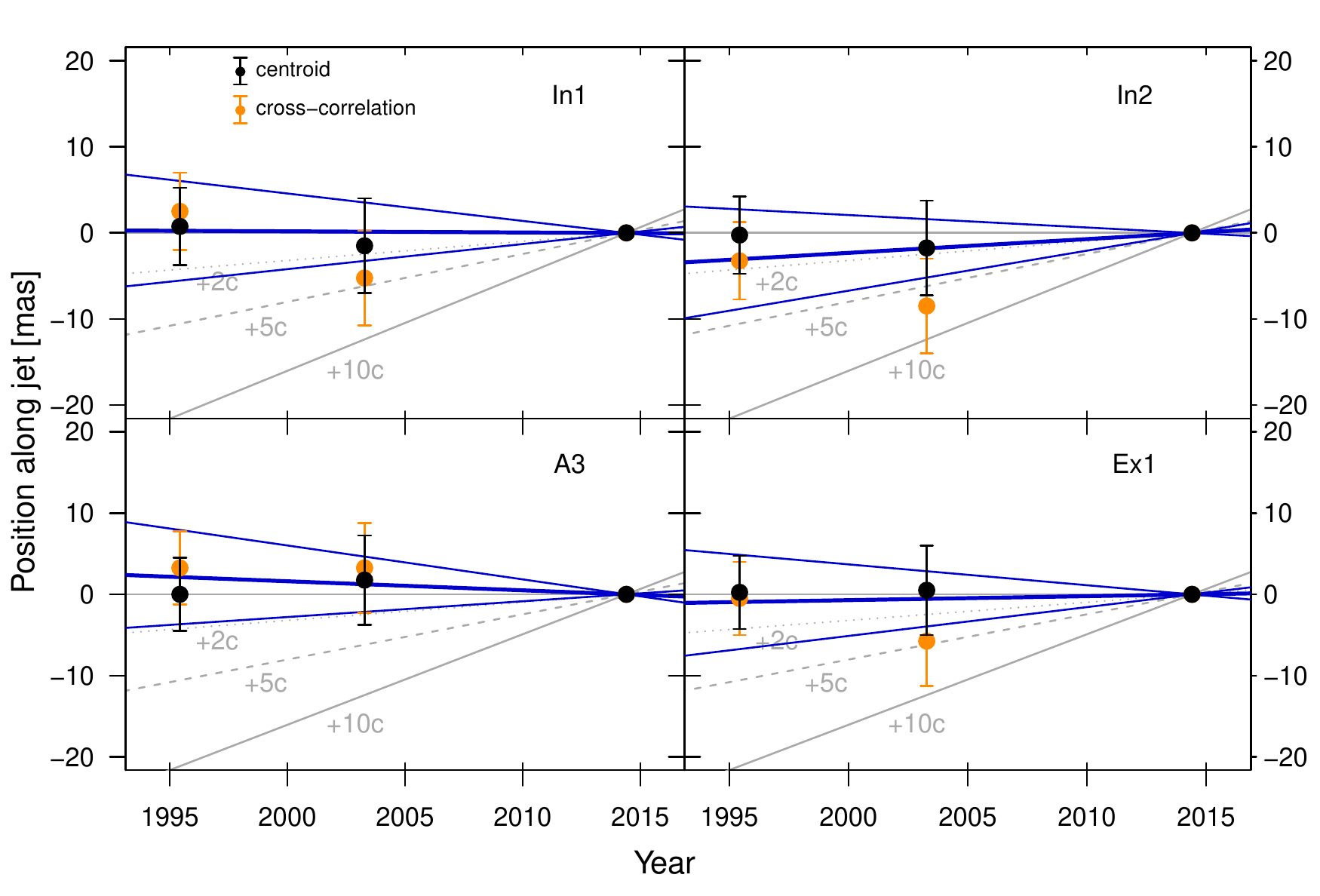}
\caption{\label{fig:speedsgals} Shift of 4 nearby galaxies to the jet
  (name noted at upper right of each panel) versus time relative to
  the 2014 measured position.  Black points represent the
  contour-derived shifts and orange points the cross-correlation
  derived shifts. As a guide, lines corresponding to an apparent
  forward speed of 2$c$ (dotted gray) and 5$c$ (dashed gray) and 10$c$
  (solid gray) are plotted in each subfigure. The thick solid blue
  line is the best-fit weighted linear regression model to all points,
  while the thinner blue lines show the 2$\sigma$ (95\%) upper and
  lower limit slopes.}
\end{figure*}

\renewcommand{\arraystretch}{1.5}
\begin{deluxetable*}{l|ccccc|cccc}
\tablewidth{0pt}
\tablecaption{\label{table:results} Positions Relative to 2014 (Parallel and Transverse to Jet Axis)\tablenotemark{a}}
\tablehead{
& 
\multicolumn{5}{c}{\underline{$\,\,\,$Contour Method$\,\,\,$}} & 
\multicolumn{4}{c}{\underline{$\,\,\,$Cross-Correlation Method$\,\,\,$}}\\
Name &
R$_{ap}$ & $\delta x_\mathrm{1995}$ & $\delta y_\mathrm{1995}$ & $\delta x_\mathrm{2003}$ & $\delta y_\mathrm{2003}$ &
$\delta x_\mathrm{1995}$ & $\delta y_\mathrm{1995}$ & $\delta x_\mathrm{2003}$ & $\delta y_\mathrm{2003}$ \\
(1) & (2) & (3) & (4) & (5) & (6) & (7) & (8) & (9) & (10) 
}
\startdata
A1  &  &\phantom{$-$}0.16$\pm$ 0.19  &$-$0.07$\pm$ 0.19  &$-$0.04$\pm$ 0.23  &$-$0.18$\pm$ 0.22  &\phantom{$-$}0.21$\pm$ 0.18  &$-$0.04$\pm$ 0.18  &$-$0.03$\pm$ 0.22  &\phantom{$-$}0.04$\pm$ 0.22  \\
A2  &10  &\phantom{$-$}0.07$\pm$ 0.18  &\phantom{$-$}0.07$\pm$ 0.18  &$-$0.02$\pm$ 0.22  &\phantom{$-$}0.03$\pm$ 0.22  &\phantom{$-$}0.41$\pm$ 0.18  &\phantom{$-$}0.26$\pm$ 0.18  &\phantom{$-$}0.04$\pm$ 0.22  &\phantom{$-$}0.08$\pm$ 0.22  \\
B1  &  &\phantom{$-$}0.27$\pm$ 0.20  &$-$0.20$\pm$ 0.20  &$-$0.18$\pm$ 0.23  &$-$0.07$\pm$ 0.23  &\phantom{$-$}0.31$\pm$ 0.18  &$-$0.11$\pm$ 0.18  &$-$0.22$\pm$ 0.22  &$-$0.09$\pm$ 0.22  \\
B2  &6.2  &\phantom{$-$}0.04$\pm$ 0.18  &\phantom{$-$}0.04$\pm$ 0.18  &$-$0.08$\pm$ 0.22  &\phantom{$-$}0.05$\pm$ 0.22  &\phantom{$-$}0.18$\pm$ 0.18  &\phantom{$-$}0.10$\pm$ 0.18  &$-$0.48$\pm$ 0.22  &\phantom{$-$}0.37$\pm$ 0.22  \\
C1  &8.2  &\phantom{$-$}0.04$\pm$ 0.18  &\phantom{$-$}0.00$\pm$ 0.18  &$-$0.08$\pm$ 0.22  &$-$0.00$\pm$ 0.22  &$-$0.01$\pm$ 0.18  &\phantom{$-$}0.02$\pm$ 0.18  &$-$0.49$\pm$ 0.22  &\phantom{$-$}0.04$\pm$ 0.22  \\
C2  &7  &\phantom{$-$}0.01$\pm$ 0.18  &$-$0.00$\pm$ 0.18  &$-$0.02$\pm$ 0.22  &$-$0.01$\pm$ 0.22  &\phantom{$-$}0.20$\pm$ 0.18  &$-$0.07$\pm$ 0.18  &$-$0.25$\pm$ 0.22  &\phantom{$-$}0.03$\pm$ 0.22  \\
C3  &7  &$-$0.06$\pm$ 0.18  &\phantom{$-$}0.00$\pm$ 0.18  &$-$0.04$\pm$ 0.22  &\phantom{$-$}0.07$\pm$ 0.22  &\phantom{$-$}0.01$\pm$ 0.18  &$-$0.08$\pm$ 0.18  &$-$0.15$\pm$ 0.22  &$-$0.03$\pm$ 0.22  \\
D  &  &$-$0.14$\pm$ 0.19  &$-$0.23$\pm$ 0.19  &\phantom{$-$}0.15$\pm$ 0.23  &$-$0.02$\pm$ 0.23  &\phantom{$-$}0.03$\pm$ 0.18  &$-$0.18$\pm$ 0.18  &\phantom{$-$}0.02$\pm$ 0.22  &$-$0.18$\pm$ 0.22  \\
H2  &5  &$-$0.01$\pm$ 0.18  &$-$0.01$\pm$ 0.18  &$-$0.04$\pm$ 0.22  &$-$0.09$\pm$ 0.22  &$-$0.08$\pm$ 0.18  &$-$0.05$\pm$ 0.18  &$-$0.05$\pm$ 0.22  &$-$0.29$\pm$ 0.22  \\
H3  &  &$-$0.13$\pm$ 0.19  &$-$0.01$\pm$ 0.19  &\phantom{$-$}0.17$\pm$ 0.26  &$-$0.38$\pm$ 0.26  &\phantom{$-$}0.05$\pm$ 0.18  &$-$0.13$\pm$ 0.18  &$-$0.06$\pm$ 0.22  &$-$0.00$\pm$ 0.22  
\enddata
\tablenotetext{a}{Units of columns 2$-$10 are in reference frame pixels (25 mas)}
\end{deluxetable*}

\renewcommand{\arraystretch}{1.5}
\begin{deluxetable*}{lccccccccc}
\centering
\tablewidth{0pt}
\tablecaption{\label{tab:speeds} Proper Motion Measurements for 3C 273 and Field Sources}
\tablehead{
\colhead{Name} & 
\colhead{Flux} &
\colhead{Aperture} &
\colhead{SB} &
\colhead{$\mu_{app,X}$} & 
\colhead{$\mu_{app,Y}$} & 
\colhead{$\beta_{app,X}$} &
\colhead{$\beta_{app,Y}$} &
\colhead{P($\beta_{app,X}$)$>0$} &
\colhead{99\% UL $\beta_{app,X}$ } \\
 & ($\mu$Jy) & (pixels) & $\mu$Jy/$''^2$ & (mas yr$^{-1}$) & (mas yr$^{-1}$) &   &   &   
}
\startdata
A1	& 3.7	& 51x20	& 24.8	&           $-$0.19$\pm$0.16	& \phantom{$-$}0.09$\pm$0.16	&           $-$1.7$\pm$1.4	& \phantom{$-$}0.8$\pm$1.4	& \phantom{$>$}12\%	& 1.6 \\ 
A2	& 1.2	& 12.7	&  6.7	&           $-$0.07$\pm$0.22	&           $-$0.09$\pm$0.22	&           $-$0.6$\pm$1.9	&           $-$0.8$\pm$1.9	& \phantom{$>$}38\%	& 3.8 \\ 
B1	& 2.5	& 14.6	& 17.2	&           $-$0.22$\pm$0.16	& \phantom{$-$}0.19$\pm$0.16	&           $-$2.0$\pm$1.4	& \phantom{$-$}1.7$\pm$1.4	& \phantom{$>$} 8\%	& 1.3 \\ 
B2	& 1.0	& 9	&  9.0	&           $-$0.01$\pm$0.22	&           $-$0.06$\pm$0.22	&           $-$0.1$\pm$1.9	&           $-$0.6$\pm$1.9	& \phantom{$>$}48\%	& 4.3 \\ 
C1	& 3.0	& 19.8	& 10.7	& \phantom{$-$}0.10$\pm$0.15	&           $-$0.02$\pm$0.15	& \phantom{$-$}0.9$\pm$1.4	&           $-$0.2$\pm$1.4	& \phantom{$>$}75\%	& 4.2 \\ 
C2	& 0.8	& 7.3	& 10.6	&           $-$0.06$\pm$0.15	& \phantom{$-$}0.03$\pm$0.15	&           $-$0.5$\pm$1.4	& \phantom{$-$}0.3$\pm$1.4	& \phantom{$>$}36\%	& 2.8 \\ 
C3	& 1.6	& 10.5	& 10.6	& \phantom{$-$}0.07$\pm$0.15	& \phantom{$-$}0.03$\pm$0.15	& \phantom{$-$}0.6$\pm$1.4	& \phantom{$-$}0.3$\pm$1.4	& \phantom{$>$}67\%	& 3.9 \\ 
D	& 1.9	& 10.1	& 17.6	& \phantom{$-$}0.02$\pm$0.16	& \phantom{$-$}0.26$\pm$0.16	& \phantom{$-$}0.2$\pm$1.4	& \phantom{$-$}2.4$\pm$1.4	& \phantom{$>$}55\%	& 3.5 \\ 
H3	& 4.3	& 22x44	& 18.0	& \phantom{$-$}0.02$\pm$0.16	& \phantom{$-$}0.14$\pm$0.16	& \phantom{$-$}0.2$\pm$1.4	& \phantom{$-$}1.3$\pm$1.4	& \phantom{$>$}56\%	& 3.5 \\ 
H2	& 0.5	& 8.7	&  6.1	& \phantom{$-$}0.07$\pm$0.15	& \phantom{$-$}0.11$\pm$0.15	& \phantom{$-$}0.6$\pm$1.4	& \phantom{$-$}1.0$\pm$1.4	& \phantom{$>$}67\%	& 3.9 \\
\hline
In1*	& 0.0	& 	&  0.0	&           $-$0.01$\pm$0.15	&           $-$0.06$\pm$0.15	&           $-$0.1$\pm$1.4	&           $-$0.6$\pm$1.4	& \phantom{$>$}46\%	& 3.2 \\ 
In2*	& 0.0	& 	&  0.0	& \phantom{$-$}0.16$\pm$0.15	& \phantom{$-$}0.02$\pm$0.15	& \phantom{$-$}1.4$\pm$1.4	& \phantom{$-$}0.1$\pm$1.4	& \phantom{$>$}85\%	& 4.7 \\ 
A3*	& 0.0	& 	&  0.0	&           $-$0.11$\pm$0.15	& \phantom{$-$}0.27$\pm$0.15	&           $-$1.0$\pm$1.4	& \phantom{$-$}2.4$\pm$1.4	& \phantom{$>$}23\%	& 2.3 \\ 
Ex1*	& 0.0	& 	&  0.0	& \phantom{$-$}0.05$\pm$0.15	& \phantom{$-$}0.06$\pm$0.15	& \phantom{$-$}0.4$\pm$1.4	& \phantom{$-$}0.5$\pm$1.4	& \phantom{$>$}63\%	& 3.7 \\ 
fs\tablenotemark{$\dagger$}	& 0.0	& 	&  0.0	& \phantom{$-$}1.79$\pm$0.39	& \phantom{$-$}0.65$\pm$0.37	& \phantom{$-$}16.1$\pm$3.5	& \phantom{$-$}5.8$\pm$3.3	& \phantom{$>$}100\%	& 24.3 
\enddata
\tablenotetext{*}{Background Galaxy}
\tablenotetext{$\dagger$}{Foreground Star}
\tablenotetext{}{\emph{Note--}$\beta_{app}$ upper limit values are calculated for the background galaxies only as a convenient comparison to the values of the jet knots.  These values are not physically meaningful.}
\label{table:speeds}
\end{deluxetable*}

\begin{figure*}[!t]
\centering
\includegraphics[width=5.35in]{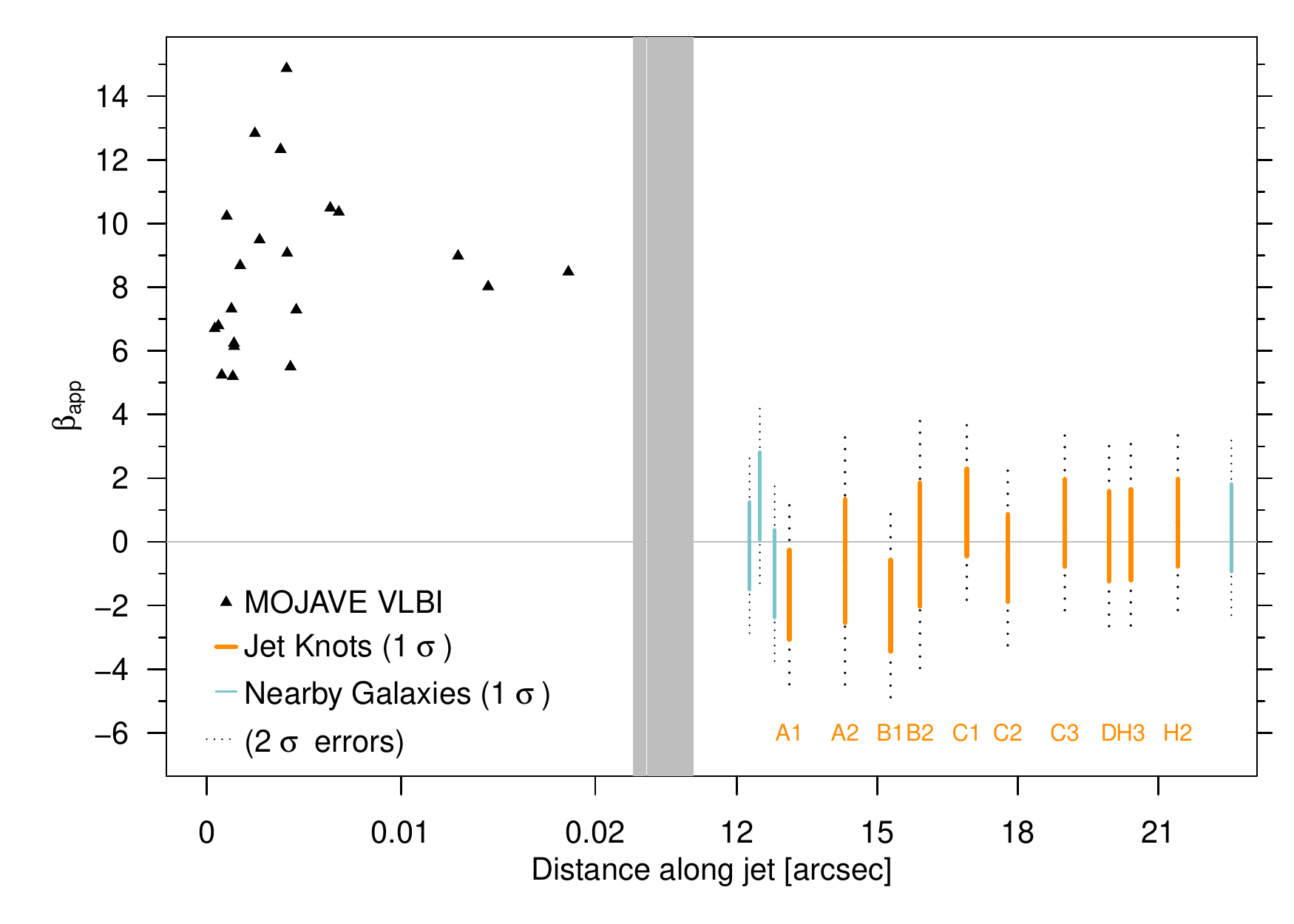}
\caption{\label{fig:speedsVSdist} A comparison of jet speeds with
  distance in the 3C~273 jet.  The radio VLBI-measured jet speeds for
  the sub-arcsecond scale jet from \cite{lister2013} are shown as
  black triangles (note that the error bars are smaller than the symbol
  size). The distance scale is broken at the gray bar to allow the
  near-and-far velocity fields to be compared.  Our data for the
  kpc-scale jet of 3C~273 is shown as orange lines spanning the
  1$\sigma$ error range, with dotted black lines giving the 2$\sigma$
  error range.  The blue lines and dotted extensions give the 1$\sigma$
  and 2$\sigma$ measurements for the four background galaxies labeled
  in Figure~\ref{fig:referencesys}.}
\end{figure*}

\section{Discussion}
\label{sec:discussion}

\subsection{The velocity of the knots in the kpc-scale Jet}
As shown in Table~\ref{table:speeds}, we do not detect significant
proper motions in any of the knots in the jet of 3C~273. Only for the bright knots A1 and A2 is there a slight case for a significant
negative proper motion, just above the 1$\sigma$ level, but we do not
claim this as a robust detection. These knots, like all other knots, do
not show any significant flux change over the 20 year timespan of the
study, and an examination of the isophotes used in the contour method
of position measurement did not suggest that any major change in knot
shape (such as an increased north-west extension of the knot) could be
responsible for the observed negative value.
As shown in column 9 of Table~\ref{table:speeds}, none of the knots
has a probability of speed greater than zero which rises to the level
of significance (i.e, $>$95\%). In general, the lack of knot
proper-motion detections is not due to lack of proper-motions
sensitivity in our study: if the knots in 3C273 had motions on the
order of 5-7$c$ (corresponding to 0.56$-$0.78 mas yr$^{-1}$), as found
previously in M87 and 3C264, we would have been able to detect these
motions. The sensitivity of the study is also demonstrated by the
significant proper motion measured for the foreground star (`fs' in
Table~\ref{table:speeds}).

We show in Figure~\ref{fig:speedsVSdist} a comparison of our kpc-scale
proper motion measurements with the parsec-scale jet speeds probed by
radio interferometry by the MOJAVE project \citep{lister2013}. The
independent axis is distance measured from the core along the jet
direction. The black triangles are the VLBI jet speeds (error bars are
less than the symbol size), which reach values up to 15$c$. Our
results for the kpc-scale knots are shown as orange lines, spanning
1$\sigma$ errors, with dotted-black-line extensions representing the
2$\sigma$ error range. The distance scale is linear but with a break
to show the two datasets side-by-side. We also show for comparison in
blue our measurements for the four nearby galaxies labeled in
Figure~\ref{fig:referencesys}. These data points counter the slight
impression that there is a bias towards more positive values of the
proper motions with increasing distance along the jet, ruling out that
this is due to any systematic bias in the image registration. Indeed,
the range of speeds observed for the background galaxies, known to
have a proper motion of absolutely zero, suggests that the spread in
knot speeds is due to the random measurement error.

The VLBI speed data suggest the possibility of a deceleration with
distance already starting on parsec-scales, as shown in
Figure~\ref{fig:speedsVSdist}. If the three most distant points
measured by VLBI accurately represent the maximum speed compared to
the highest upstream speed of nearly 15$c$, both exponential and
linear fits suggest the jet will reach mildly relativistic speeds
($\beta_{app}\approx 1$) within an arcsecond (2.6 kpc, projected) of
the core, well before the distance to the optical jet which begins
$\approx$12$''$ further on from the core. A similar result is seen in
M87, where the maximum speed reached at HST-1 of 6$c$ drops to speeds
of $<$2$c$ within a kpc \citep{biretta1999,meyer2013_m87}. Indeed, it
should be emphasized that while we refer to the jets of M87, 3C~264,
and 3C~273 as ``kpc-scale'', the jet of 3C~273, at $\approx$200-400
kpc (deprojected) is likely 40-100 times longer than the lower-power
FR I sources.  A recent acceleration study by the MOJAVE program
\citep{homan2015} shows that some sources switch from
acceleration to deceleration beyond sub-kpc
scales. It may simply be that even though 3C~273 starts out with a much
higher speed, the distance to knot A is such that the jet has slowed
down appreciably.  It is also worth noting that 3C~273 is somewhat
unusual in not having hotspots, which are usually interpreted as the
point of final deceleration of a powerful, still-relativistic jet,
which may also indicate that the jet has slowed either before or at
the optical jet.

\subsection{Constraints on the Physical Conditions in the kpc-scale Knots}
\begin{figure}[!t]
\centering
\includegraphics[width=3.5in]{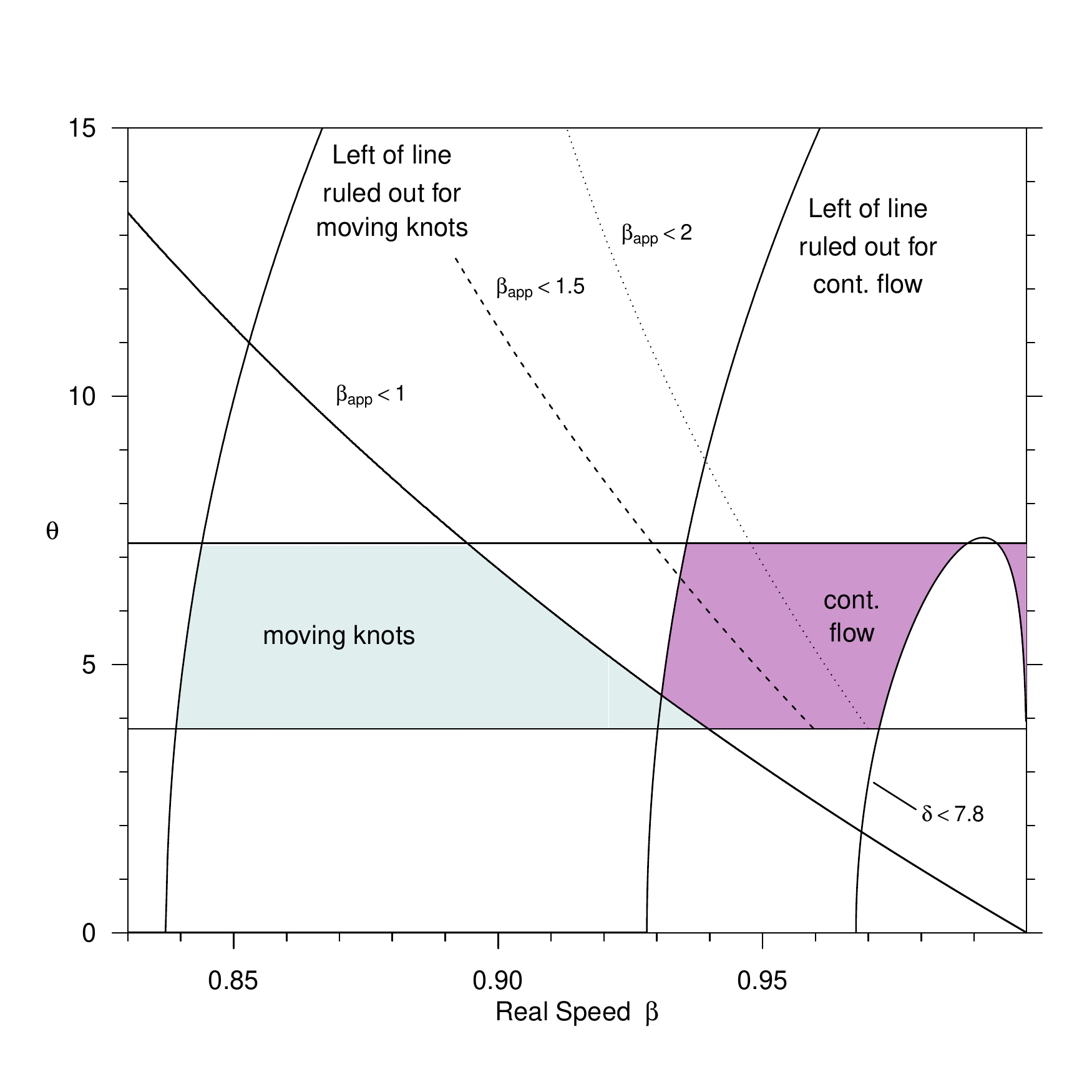}
\caption{\label{fig:betatheta} This plot shows the allowed range of
  real jet speed $\beta$ and the angle to the line-of-sight $\theta$
  under two possible scenarios.  The light-blue shaded area
  corresponds to the allowed region if the optical knots are moving
  components, while the mauve region corresponds to the allowed
  parameters if the optical jet is a continuous flow (and the knots
  are shock features). For both areas, the left boundary is formed by
  the curves dictated by the jet-to-counterjet ratio $R>10^4$ in each
  case (\citealt{conway1993}, see also \citealt{geo06}), and the
  entire jet is subject to the limits on viewing angle of
  3.8$^\circ<\theta<$7.2$^\circ$ dictated by the condition that the
  jet not exceed 1 Mpc in length or a $\Gamma>50$ in the parsec-scale
  jet (see text). For the moving knots case, the thick upper-right
  boundary corresponds to the limit from our observations that
  $\beta_{app}<1c$ for knot A1. In the continuous flow case, there is
  an additional boundary from the condition that $\delta<$7.8 from
  \cite{mey15_0637}. Within the allowed `moving knots' region, the
  maximum speed $\beta=0.94$, which corresponds to a limit
  $\Gamma\leqslant 2.9$ and $\delta\leqslant5.5$ at this point.}
\end{figure}

We now discuss the limits the present and previous observations yield
for two important physical parameters: the real speed $\beta$ and the
angle to the line-of-sight $\theta$
(Figure~\ref{fig:betatheta}). Understanding the allowed parameter
space will in turn allow us to evaluate the energetic requirements and
fitness of different physical models for the kpc-scale knots. Two
possible scenarios are before us: either the knots are moving packets
of plasma, or these features represent `standing shock' features which
move much more slowly than the bulk plasma speed. 

The maximum observed speed of 15$c$ measured by the MOJAVE project
implies an angle no larger than 7.6$^\circ$ as an absolute maximum
(assuming $\beta=1$) or 7.2$^\circ$ to the line-of-sight for the
parsec-scale jet if we assume that $\Gamma<50$ as implied by the
maximum speeds observed for the entire sample of VLBI-observed jets
\citep[e.g.,][]{lister2009}.  No deviations are seen from parsec to
kpc scales in 3C~273 which would suggest any bending in or out of the
line-of-sight. We therefore adopt the 7.2$^\circ$ as the maximum angle for
the kpc-scale jet as well. A further global limit on the angle to the
line-of-sight can be calculated by assuming that the 24$''$
jet does not exceed 1 Mpc in total (deprojected) length, as very few
 radio galaxies exceed this length \cite[e.g. 3C 236,][]{schilizzi2001}; this gives a
lower limit $\theta=3.8^\circ$. These minimum and maximum angles are
plotted as horizontal lines in Figure~\ref{fig:betatheta}.

If we assume that the optical knots are `ballistic' packets of moving
plasma, a limit of $\beta_\mathrm{app}<1c$ on the knot speeds can be
used to derive a limit in the $\beta-\theta$ plane, as shown in
Figure~\ref{fig:betatheta} as the slanting black line bounding the
right of the allowed zone under the moving knots scenario (more
conservative limits of 1.5$c$ and 2$c$ are shown as dashed and dotted
lines as labeled). Further, the observation that the jet-to-counterjet
ratio $R$ exceeds 10$^4$ \citep{conway1993} leads to the left boundary
to this area, where
R=(1+$\beta$cos$\theta$)$^{m+\alpha}$/(1-$\beta$cos$\theta$)$^{m+\alpha}$. In
the case of moving knots $m=3$ and for a continous flow $m=2$, while
we take $\alpha$=0.8 from radio observations \citep[see][]{geo06}.
The jet-to-counterjet ratio limit thus also leads to a left bound on
the allowed area for a continuous flow jet, as shown in
Figure~\ref{fig:betatheta}.  Note that the two zones (moving knots
versus continuous flow) are nearly mutually exclusive only under the
$\beta_\mathrm{app}<1c$ limit. Under the more conservative (larger)
upper limits for $\beta_\mathrm{app}$, the moving knots allowed zone
extends to higher $\beta$ values, overlapping with purple-shaded the
continuous flow region.

The allowed range of $\beta$ according to the boundaries in
Figure~\ref{fig:betatheta} in the moving knots case under
$\beta_\mathrm{app}<1c$ is 0.84$<\beta<0.94$, corresponding to
1.8$<\Gamma<2.9$. The maximum $\Gamma$ increases to 3.6 for a limit
  $1.5c$ and $4.1$ for a limit of $2c$. The maximum Doppler beaming factor
in the allowed zone under any of the $\beta_\mathrm{app}$ limits is
found at the point of minimum angle (3.8$^\circ$) and maximum jet
speed. For $\beta_\mathrm{app}<1c$, $1.5c$, and $2c$, respectively, the
upper limit on $\delta$ is 5.5, 6.7, and 7.6.  As we discuss below, these
$\delta$ values are considerably lower than the $\delta=20$ required (under
equipartition) in the large-scale jet if the X-ray emission from the
knots is from the IC/CMB process.


\subsection{Implications for the IC/CMB Model for the X-ray emission}
The kpc-scale jet of 3C~273 has been detected by \emph{Chandra} in the
X-rays \citep{marshall2001,jester2006}, where the hard spectrum and
high flux level of the knots shows that the X-rays are due to a
separate component from the radio-optical synchrotron spectrum.
Indeed, HST observations by \cite{jester2007} show that the spectrum
is already upturning into this second component at UV energies. The
jet of 3C~273 is one of dozens of `anomalous' X-ray jets discovered by
\emph{Chandra} to have hard and high X-ray fluxes in the knots which require a
second component \citep[e.g.,][for a review]{harris2006}.  The most
favored explanation for the X-rays in these jets has been that the
large-scale jet remains as highly relativistic as the parsec-scale
jet, with $\Gamma$$\approx$10 or more. Coupled with a small angle to
the line-of-sight, the increased Doppler boosting suggests that the
X-rays could be consistent with inverse Compton upscattering of CMB
photons by the same electron population that produces the
radio-optical synchrotron spectrum, assuming the electron energy
distribution can be extended to much lower energies than traced by GHz
radio observations \citep{tavecchio2000,celotti2001,jester2006}.  Alternatively,
it has been suggested that the second component producing the X-rays
could be synchrotron in origin, from a separate electron energy
distribution which reaches multi-TeV energies
\citep{harris2004,kataoka2005,hardcastle2006,jester2006,uchiyama2006}. The
differences between the IC/CMB and synchrotron mechanisms are important: the former requires a fast and powerful jet
(sometimes near or super-Eddington), while the latter suggests a
slower and less powerful jet on kpc scales \citep{geo06}. The main
opposition to the synchrotron interpretation remains its 
unclear origin \citep{atoyan2002,aharonian2002,liu2015}.

While IC/CMB is a popular explanation for anomalous X-ray jets, it has
now been ruled out explicitly in three cases: for PKS~1136-135 based
on UV polarization in the second spectral component \citep{cara2013},
and for PKS~0637-752 and 3C~273 based on non-detection of the
gamma-rays implied by the IC/CMB model \citep{mey14,mey15_0637}, an
idea first proposed by \cite{geo06}. In the case of 3C~273, we 
show here in an independent way that the IC/CMB model is also 
disfavored by our proper motions upper limits.

It has already been shown that the X-rays from the knots of 3C~273 and
similar jets, can only be compatible with an IC/CMB origin if the
knots are moving packets.  This is because in the case of
particle-accelerating standing shocks in the IC/CMB model, the
extremely long (hundreds of Mpc) cooling length of the low energy
X-ray emitting electrons would result in continuously-emitting X-ray
jets, instead of the observed knotty appearance
\citep{atoyan2002}. This is avoided in the case of a moving packet of
plasma, as the low energy electrons remain confined within the packet.

The minimum power configuration for the first and brightest knot A1 is
that of equipartition between radiating electron and magnetic field
energy density.  With the additional assumption that the Lorentz
factor equals the Doppler factor, $\Gamma=\delta$ , this requires
$\delta=20$. All configurations, however, with $\delta>7.6$ are
excluded because of the constraints discussed above. This requires
that we move away from the equipartition power requirement of
$10^{48}$ erg s$^{-1}$ (assuming one proton per radiating electron),
to $5 \times 10^{48}$ erg s$^{-1}$ for the minimum power configuration
in the permitted zone at the extreme edge where $\theta=3.8^\circ$,
$\delta=5.5$. Elsewhere in the allowed moving-knots zone the minimum
power is even higher.
 
We compare now this power to the Eddington luminosity of the source.
Mass estimates for the black hole of 3C~273 vary widely, from $2\times
10^7\,M_\odot$ \citep{wang2004} to $4\times 10^8\,M_\odot$
\citep{pian2005} to $6.6\times 10^9\,M_\odot$
\citep{paltani2005}. Even for the highest mass estimate, the Eddington
luminosity is $10^{48}$ erg s$^{-1}$. This is barely compatible with
the equipartition configuration, which however we disfavor because it
does not comply with our angle and superluminal motion constraints.
The minimum jet power compatible with $\delta<5.5$ is five times
higher than the Eddington luminosity, adopting the highest black hole
mass for 3C 273.
Given that the jet power is in general found to be sub-Eddington
\citep{ghisellininature} we disfavor the IC/CMB mechanism for the
production of the X-rays, as it requires a power of at least five
times Eddington, and up to several hundred times Eddington depending on
the black hole mass.

\section{Conclusions}

We have used new and archival HST V-band imaging of the optical jet in
3C~273 to look for significant proper motions of the major knots over
the 19 years between June 1995 and May of 2014. We have described a
method of image registration based on background galaxies; in the 2014
deep ACS imaging, our systematic error in the stacking is 0.44 mas,
while the systematic error of registration for the 1995 and 2003
epochs of WFPC2/PC imaging is 4.5 and 2.8 mas, respectively. We have
used both a two-dimensional cross-correlation and a centroiding
technique to measure relative shifts in the knots both along and
perpendicular to the jet direction. Our results show that all knots
have speeds consistent with zero with typical 1$\sigma$ errors on the
order of 0.1$-$0.2 mas/year or 1.5$c$, and with 99\% upper limit
values ranging from $1-5$c. We have used nearby background galaxies to
show that these limits are consistent with stationary objects in the same field.

These results suggest that the knots in the kpc-scale jet, if they are
moving packets of plasma, must be relatively slow, in
  agreement with previous studies based on jet-to-counterjet ratios in radio-loud populations \citep{arshakian2004,mullin2009}.  Assuming
  that the jet either remains at the same speed or decelerates as you
  move downstream, the $2\sigma$ upper limit speed derived from all
  knots combined of $1c$ suggests that the entire optical jet is at
  most mildly relativistic, with a maximum Lorentz factor of
  $\Gamma<2.9$. However, we cannot rule out the possibility that the
  knots are standing shock features in the flow, where the bulk plasma
  moves through the features with a higher $\Gamma$.  The best limits
  on the bulk plasma speed thus remain the limits derived from the
  non-detection of the IC/CMB component in gamma-rays by
  \cite{mey15_0637}, where $\delta<7.8$ is implied assuming
  equipartition magnetic fields.

Finally, we show that the observed upper limits on the proper motion
of the knots confirms that the a near-equipartition IC/CMB model for
the X-rays of the kpc-scale knots is ruled out. The equipartition
IC/CMB model requires that the knots are ballistic packets of moving
plasma moving at the bulk speed $\Gamma\approx15-20$ which would imply
proper motions on the order of 10$c$ or 1.12 mas/year which could have
been detected in our study; our upper limits easily rule this out at a
high level of significance ($>$5$\sigma$). Moving away from
equipartition conditions, an IC/CMB model consistent with our
observations requires a jet power on the order of five to several
hundred times the Eddington limit, and is thus energetically disfavored.

In comparison to other recent HST observations of lower-power optical
jets M87 and 3C264, where highly superluminal speeds (6$-$7$c$) have been
observed in the optical kpc-scale jet, our first proper-motion study
of a powerful quasar jet reveals no significant
proper motions.  It remains to be seen whether this is
because the jet has truly decelerated and is only mildly relativistic,
or because the knot features in sources like M87 and 3C~273 represent
very different things: moving packets of plasma in the first instance
and standing shocks in the second.
\label{sec:conclusion}

 \acknowledgments E.T.M. acknowledges HST Grant GO-13327. E.T.M. and M.G. also acknolwedge NASA grant 14-ADAP14-0122.

\bibliographystyle{apj}
\bibliography{3c273,3c273_second}


\appendix
\section{CTE Correction Maps}
\label{appendixA}
Losses due to charge transfer inefficiency (CTI=1-CTE) in the WFPC2
detectors is fairly well-studied problem. The first correction
formulae were published by \cite[][hereafter WHC99]{whitmore_cte_99},
and later updated by \cite{dolphin_cte_00}, in both cases based on
observations of stars. A comparison between the two shows reasonably
good agreement \citep{whitmore_cte_02}, with the WHC99 formulae
producing smaller corrections at very low flux levels. We have used
the WHC99 formulae in our corrections, but the method of producing
pixel-by-pixel maps presented here could be used with any set of
corrections.

Applying the WHC99 CTE loss correction formulae directly to the
measured fluxes for the jet or galaxies in our imaging would be
inappropriate because they are resolved, while the formulae are based
on fixed-aperture observations of stars. We also wished to produce
CTE-corrected frames from the pipeline-produced `c0f' files to use in
registering the images through the background galaxies, where accurate
flux levels are obviously helpful. Therefore, a pixel-by-pixel
correction `map' for each raw image is needed.  Note that this method
is not flux-conserving, but seems to work well in recovering the total
flux in bright, resolved, but relatively compact sources such as the
jet knots and background galaxies. 

To produce these maps, we first measured the background level in each
raw image -- usually about 8 counts/pixel in 1995 and 3 counts/pixel
in 2003. In order to calculate the corrected flux in each pixel, we
need the modified Julian date (MJD) of the observation, the $x$ and
$y$ location on the detector, the background flux level, the `source'
flux (total - background) and the WHC99 formulae.  We have assumed
that the CTE corrections have the same form (though not the same
parameters) when based on the pixel value as when based on the total
flux of a star within a radius=2 pixel aperture.  We transformed
between these representations of the correction by using a `known'
PSF, generated by the tinytim package, for the F622W filter using a
powerlaw form and otherwise standard parameters. Since the original
correction formulae were based on a 2-pixel-radius aperture, we only
need the PSF to be defined within an aperture of this size.  

The correction at each pixel $i$ is assumed to have the form:

\begin{equation}
x_{cts,i} = \left(\alpha_x + \beta_x \log\left(\mathrm{cts}_{0,i}\right)\right)\mathrm{cts}_{0,i}
\end{equation}

\begin{equation}
y_{cts,i} = \left(\alpha_y + \beta_y \log\left(\mathrm{cts}_{0,i}\right)\right)\mathrm{cts}_{0,i}
\end{equation}

where the counts in the pixel \emph{before CTE losses} is cts$_{0,i}$
and $x_{cts,i}$ and $y_{cts,i}$ are the corrections for CTE losses in the
x and y directions, respectively, in DN. We want to determine the values of
the $\alpha$ and $\beta$ parameters individually for each pixel.

To do that, we choose a vector of simulated observed stars with counts
\emph{before CTE losses} of $T^*_0$, and measured counts $T^*_M$ in
the two-pixel radius aperture. Starting with a vector of $T^*_M$
values, the WHC99 formulae and properties of the pixel and image are
used to get the values of $T^*_0$, $x_{cts}$ and $y_{cts}$, noting
that $T^*_0 = x_{cts} + y_{cts} + T^*_M$. These vectors can be related
to the individual pixels falling within the aperture in the following
way:

\begin{equation}
y_{cts} = \sum_i y_{cts,i}f_i
\end{equation}

where $y_{cts}$ is the total correction due to y-direction CTE losses
for all the pixels within the aperture.  Because we have used a
circular aperture, the fraction $f_i$ is used to only count the
portion of the pixel that falls within the two-pixel radius, which we
have assumed is centered on the PSF. A similar equation holds in the
x-direction. The vectors based on `stars' can be used to derive the
$\alpha$/$\beta$ parameters from the following linear relation:

\begin{equation}
\frac{y_{cts}}{S_1} = \alpha_y + \beta_y\frac{S_2}{S_1}
\label{lineareq}
\end{equation}

where

\begin{equation}
S_1 = \sum_i \mathrm{cts}_{0,i}f_i = T^*_0\sum_i c_i f_i = T^*_0
\end{equation}

\begin{equation}
S_2 = \sum_i \mathrm{cts}_{0,i}\log\left(\mathrm{cts}_{0,i}\right)f_i = T^*_0\sum_i c_i f_i\log\left(T^*_0\mathrm{cts}_{0,i}\right)
\end{equation}

Note that the quantities $c_i$ define the PSF inside the aperture such that

\begin{equation}
\sum_i c_i f_i = 1.
\end{equation}

Using values of $T_M$ around the value of the observed pixel counts, a
linear least-squares fitting can easily derive the values of
$\alpha_y$ and $\beta_y$ for equation~\ref{lineareq}, and similarly
for $\alpha_x$ and $\beta_x$.  These then become the parameters for
calculating the pixel-based, rather than aperture-based correction.

\end{document}